\documentclass[acmsmall,screen,authorversion,nonacm]{acmart}

\usepackage{tabularx}
\usepackage{bm}
\usepackage{makecell}
\usepackage{subcaption}
\usepackage{balance}

\def\BibTeX{{\rm B\kern-.05em{\sc i\kern-.025em b}\kern-.08emT\kern-.1667em\lower.7ex\hbox{E}\kern-.125emX}}

\begin{document}

%
\title{Boosting the Rating Prediction with Click Data}

%
\author{ThaiBinh Nguyen}
\affiliation{%
  \institution{SOKENDAI (The Graduate University for Advanced Studies)}
  \city{Kanagawa}
  \country{Japan}}
\email{binh@nii.ac.jp}

\author{Atsuhiro Takasu}
\affiliation{%
  \institution{National Institute of Informatics}
  \city{Tokyo}
  \country{Japan}
}
\email{takasu@nii.ac.jp}

%

%
\begin{abstract}
Matrix factorization (MF) is one of the most efficient methods for rating predictions. MF learns user and item representations by factorizing the user-item rating matrix. Further, textual contents are integrated to conventional MF to address the cold-start problem. However, the textual contents do not reflect all aspects of the items. In this paper, we propose a model that leverages the information hidden in the item \textit{co-click} (i.e., items that are often clicked together by a user) into learning item representations. We develop TCF (Textual Co-Factorization) that learns the user and item representations jointly from the user-item matrix, textual contents and item co-click matrix built from click data. Item co-click information captures the relationships between items which are not captured via textual contents. The experiments on two real-world datasets (\textit{MovieTweetings}, and \textit{Bookcrossing}) demonstrate that our method outperforms competing methods in terms of rating prediction. Further, we show that the proposed model can learn effective item representations by comparing with state-of-the-art methods in classification task which uses the item representations as input vectors.
\end{abstract}

%
%

\ccsdesc[500]{Computer systems organization~Embedded systems}
\ccsdesc[300]{Computer systems organization~Redundancy}
\ccsdesc{Computer systems organization~Robotics}
\ccsdesc[100]{Networks~Network reliability}

%
\keywords{recommender system, item embedding, implicit feedback, explicit feedback, matrix factorization}

%

%
\maketitle

\section{Introduction}
Nowadays, a recommender system (RS) has become a core component of many online services. RS analyzes users' behavior and provides them with personalized recommendations for products or services that meet their needs. For example, Amazon recommends products to users based on their shopping histories; an online newspaper recommends articles to users based on what they have read. Generally, an RS can be classified into two categories: Content-based approach and Collaborative Filtering (CF)-based approach. The content-based approach creates a description for each item and builds a profile for each user's preference. In other words, the content-based approach recommends the items that are similar to items that interested the user. In contrast, CF-based approach \cite{sarwar01itembased,reference/sp/NingDK15,koren2008factorization,salakhutdinov2008a,journalscacmKoren10} relies on the behavior of each user in the past, such as users' ratings on items. The CF-based approach is domain-independent and does not require content collection as well as content analysis.

Basically, the data for a CF-based algorithm comes in the form of a rating matrix whose entries are observed ratings to items given by users. This kind of feedback is referred to as \textit{explicit feedback}. Given these observed ratings, a typical task of CF-based algorithms is to predict the unseen ratings. One of the most efficient ways to perform CF is matrix factorization (MF) \cite{koren2008factorization,salakhutdinov2008a,journalscacmKoren10} which decompose the rating matrix into latent vectors that represent \textit{users' preferences} and \textit{items' attributes}. These latent vectors are then used to predict the unseen ratings. Usually, MF-based algorithms suffer from the sparseness of the rating matrix: if a user or an item has a very few numbers of ratings, it is difficult to find a ``right'' latent vector for the user or the item; in an extreme case, if rating data is not available for an item, MF-based algorithms cannot find a latent vector for it (the \textit{cold-start} problem). To address the cold-start problem, auxiliary information such as textual contents are exploited \cite{wang2011collaborative,conf_kdd_WangWY15,conf_nips_GopalanCB14,Li_CVAE_2017}. In these models, the item representations can be inferred directly from the textual contents, even if there are no prior ratings associated with the items.

Although the textual content is a significant information for modeling item representations, it is not the only aspect that effects to the user preferences. For example, a researcher in biology may have interest in a machine learning paper although the content of a biology paper and that of a machine learning paper are different. Such item-item relationships are not captured by textual contents. In this paper, we take into account the item-item relationships by exploring an alternative which models item relationships via item co-click data. We assume that items that are clicked in the same context are similar. This is equivalent to word embedding \cite{confnipsMikolovSCCD13,pennington2014glove}, where words that appear in the same context have the similar meaning.

We propose TCF (Textual Co-Factorization), a model that leverages the co-click information for extracting the item relationships that are not captured by MF as well as textual contents. Co-click data encode not only the textual information but also the relationships between items that are frequently clicked together. Thus, integrating co-click information can help to learn effective item representations. Technically, first, we build an item-item matrix according to co-click information. TCF will simultaneously factorize the user-item matrix and item-item matrix in a shared latent space while integrating textual contents.

To sum up, we focus on the rating prediction problem, particularly for sparse data, where the number of ratings is not enough for learning good item representations. This paper is an extension of our previous work \cite{nguyen2017probabilistic} with a thorough investigation and analysis. In \cite{nguyen2017probabilistic}, we proposed EMB-MF, a model that utilizes the click data in the rating prediction task. By utilizing the click data, the EMB-MF partly solves the cold-start problem: it can predict the rating for an item that has no prior ratings as long as this item has some clicks. However, the drawback of EMB-MF is that it cannot predict the ratings for items which does not have any clicks. The proposed model, TCF, address this issue by exploiting the textual contents of the items. We demonstrate the advantage of TCF by comparing it with EMB-MF and other state-of-the-art models in rating prediction using textual contents: collaborative topic regression (CTR) \cite{wang2011collaborative}, collaborative deep learning model (CDL) \cite{conf_kdd_WangWY15}, and collaborative variational auto-encoder (CVAE) \cite{Li_CVAE_2017}.

The main contributions of this work can be summarized as follows.
\begin{itemize}
\item We proposed a probabilistic model that learns the item embedding from click data which capture the relationships between items.
\item We proposed TCF, a probabilistic model for learning item representations jointly from the rating data, the textual contents and the click data. TCF is a joint model of the stacked denoising autoencoder (SDAE) for textual contents, matrix factorization for rating data, and item embedding for click data.
\item Our extensive experiments on real-world datasets demonstrate that TCF significantly outperforms state-of-the-art methods for rating prediction, particularly for extremely sparse datasets.
\end{itemize}

The rest of this paper is organized as follows. In Section \ref{sec:preliminary} we formulate the problem and represent the background knowledge related to the method. Section \ref{sec:methodology} represents our idea in modeling items using implicit feedback and describes the probabilistic model for integrating implicit and explicit feedback data in a unified model. In Section \ref{sec:empirical_study}, we present the effectiveness of our method by comparing with state-of-the-art techniques using three public datasets. We show some related work in Section \ref{sec:related_work} and summarize the proposed model in Section \ref{sec:conclusion}.

\section{Preliminary}
\label{sec:preliminary}
\subsection{Notation and Problem Formulation}
Let us establish some notations. We use $u$ to denote a user and $i$ or $j$ to denote an item. Each observation of explicit feedback is represented by a triplet $(u, i, r_{ui})$ where $r_{ui}$ is the rating that user $u$ gave to item $i$. The explicit feedback can be represented by a matrix $R\in\mathbb{R}^{N\times M}$ where $N$ is the number of users and $M$ is the number of items. Each entry $r_{ui}$ of the matrix $R$ is either the rating of item $i$ given by user $u$ or zero if the rating is not observed (missing entries). We use $\mathcal{R}$ to denote the set of $(u,i)$-pair that $r_{ui}>0$, $\mathcal{R}_{u}$ to denote the set of items that user $u$ gave ratings, and $\mathcal{R}_i$ to denote the set of users that gave ratings to item $i$.

The implicit feedback of a user-item pair is represented by a triplet $(u, i, p_{ui})$ where $p_{ui}=1$ if implicit feedback of user $u$ to item $i$ is observed (e.g., the click, views or purchase of item $i$ by user $u$), and $p_{ui}=0$ if the implicit feedback is not observed. Implicit feedback is represented by matrix $P\in\{0,1\}^{N\times M}$.

Table \ref{tab:notations} lists the notations used throughout this paper.

\begin{table}[!t]
  \centering
  \caption{The notations used throughout the paper.}
  \label{tab:notations}
  \renewcommand{\arraystretch}{1.2}
  \begin{tabularx}{\linewidth}{c|X}
    Notation&Meaning\\
    \hline
    $N, M$ & the number users and items, respectively\\
    $\mathbf{R}$ & the rating matrix ($\mathbf{R}\in \mathbb{R}^{N\times M}$)\\
    $\mathbf{P}$ & the click matrix ($\mathbf{P}\in \{0,1\}^{N\times M}$)\\
    $\mathbf{S}$ & the positive point-wise information (PPMI) matrix ($\mathbf{S}\in \mathbb{R}^{M\times M}$)\\
    $\mathcal{R}$ & the set of $(u,i)$ pair that $r_{ui}>0$ (i.e., observed ratings)\\
    $r_{ui}$ & the rating that user $u$ gave to item $i$. This is the element of row $u$ and column $i$ of the rating matrix $\mathbf{R}$\\
    $s_{ij}$ & the element at row $i$ and column $i$ of matrix $\mathbf{S}$\\
    $\mathcal{S}$ & the set of $(i,j)$-pair that $s_{ij}>0$\\
    $K$ & the dimensionality of the latent space\\
    $\bm{\theta}_u$ & the user feature vector of user $u$ ($\bm{\theta}_u\in \mathbb{R}^{K}$)\\
    $\bm{\beta}_i$ & the item feature vector of item $i$ ($\bm{\beta}_i\in \mathbb{R}^{K}$)\\
    $\bm{\alpha}_i$ & the item context vector of item $i$ ($\bm{\alpha}_i\in \mathbb{R}^{K}$)\\
    $\bm{\theta,\beta,\alpha}$ & matrices whose columns are user feature vectors, item feature vectors, item context vectors, respectively. $\bm{\theta}=\theta_{1:N}$, $\bm{\beta}=\beta_{1:M}$, $\bm{\alpha}=\alpha_{1:M}$\\
    $\bm{\Omega}$ & the set of all model parameters\\
    $\sigma^2_R$ & the variance of the ratings\\
    $\sigma^2_\theta$ & the variance for user latent factor vectors\\
    $\sigma^2_\beta$ & the variance for the item feature vectors\\
    $\sigma^2_\alpha$ &  the variance for the item context vectors\\
    $L$ & the number of layers of the stacked denoising auto-encoder (SDAE)\\
    $\mathbf{W}_l$ & the weight parameters of layer $l$ of the SDAE\\
    $\mathbf{b}_l$ & the bias parameters of layer $l$ of the SDAE\\
    $\mathbf{X}_l$ & the output of the $l^{th}$ layer of the SDAE\\
    $\mathbf{X}_0$ & the noise corrupted matrix (the input of the SDAE)\\
    $\mathbf{X}_c$ & the original clean version of the texts (the output of the SDAE)\\
    $\sigma^2_W$ & the variance of the weight parameters of the SDAE\\
    $\mathbf{W}$ & the set of all the weight parameters of the SDAE's layers: $\mathbf{W}=[\mathbf{W}_1, \mathbf{W}_2, \dots, \mathbf{W}_L$\\
    $\mathbf{B}$ & the set of all bias parameters of the SDAE's layers: $\mathbf{B}=[\mathbf{b}_1, \mathbf{b}_2, \dots, \mathbf{b}_L]$\\
    $\sigma^2_X$ & the variance of the clean bag-of-words vectors\\
    $\lambda$ & the balance parameter between rating data and click data\\
\end{tabularx}
\end{table}

\subsection{Probabilistic Matrix Factorization}
\label{subsub:pmf}
Given rating matrix $\mathbf{R}$, probabilistic matrix factorization (PMF) \cite{salakhutdinov2008a} decomposes $\mathbf{R}$ to two low-rank matrices $\bm{\theta}$ and $\bm{\beta}$ whose columns are latent factor vectors of users and items, respectively. $\mathbf{R}=\bm{\theta}^\top\bm{\beta}$. PMF assumes that the rating $r_{ui}$ can be modeled by a normal distribution as follows.
\begin{equation}
  \centering
  P(r_{ui}|\bm{\theta}_u, \bm{\beta}_i, \sigma_R^2) = \mathcal{N}(\bm{\theta}_u^\top\bm{\beta}_i, \sigma_R^{2})
\end{equation}
where $\mathcal{N}(.)$ is a normal distribution; $\bm{\theta}_u^\top\bm{\beta}_i$ and $\sigma_R^{2}$ are its expectation and variance, respectively.

The likelihood function for the entire matrix $\mathbf{R}$ is as follows:
\begin{equation}
p(\mathbf{R}|\bm{\theta}, \bm{\beta}, \sigma_R^2) = \prod_{(u,i)\in\mathcal{R}}p(r_{ui}|\bm{\theta}_u,\bm{\beta}_i, \sigma^2_R)
\end{equation}

Further, the zero-mean sphere Gaussian distributions are placed on the latent factor vectors of users and items as follows.
\begin{equation}
\begin{aligned}
p(\bm{\theta}|\sigma_\theta^2) &= \prod_u\mathcal{N}(\bm{\theta}_u|\mathbf{0}, \sigma_\theta^2\mathbf{I}_K)\\
p(\bm{\beta}|\sigma_\beta^2) &= \prod_i\mathcal{N}(\bm{\beta}_i|\mathbf{0}, \sigma_\beta^2\mathbf{I}_K)
\end{aligned}
\end{equation}

The posterior distribution is as follows.
\begin{equation}
\begin{aligned}
p(\bm{\theta}, \bm{\beta}|\mathbf{R}, \sigma_R^2)&\propto p(\mathbf{R}|\bm{\theta}, \bm{\beta}, \sigma_R^2)p(\bm{\theta}|\sigma_\theta^2)p(\bm{\beta}|\sigma_\beta^2)\\
\end{aligned}
\end{equation}

The model parameters are learned by maximizing the likelihood function, which is equivalent to minimizing the negative log posterior distribution which is given in Eq.\ref{eq:pmf_objective_function}.
\begin{equation}
      \label{eq:pmf_objective_function}
     \begin{aligned}
    \mathcal{{L}}(\bm{\Omega}) & =\frac{1}{2}\sum_{(u,i)\in \mathcal{R}}[r_{ui}-(\bm{\theta}_u^\top\bm{\beta}_i)]^2 +\frac{\lambda_\theta}{2}\sum_{u=1}^{N}||\bm{\theta}_u||^2+\frac{\lambda_\beta}{2}\sum_{i=1}^{M}||\bm{\beta}_i||^2
    \end{aligned}
\end{equation}
where $\lambda_\theta=\sigma_R^2/\sigma_\theta^2$, and $\lambda_\beta=\sigma_R^2/\sigma_\beta^2$; $||.||^2$ is the $L2$-norm of a vector.

The objective function in Eq.\ref{eq:pmf_objective_function} can be minimized using stochastic gradient descent (SGD) or alternating least square (ALS) method.

\subsection{Textual Embedding}
We study how to apply an unsupervised deep learning-based model for learning item representations from textual data. We adopt the stacked denoising auto-encoder (SDAE) \cite{conf_kdd_WangWY15} as the textual representation learning method because of its advantages in modeling short texts and non-linear relations.

\begin{figure}
    \centering
        \includegraphics[width=0.4\textwidth]{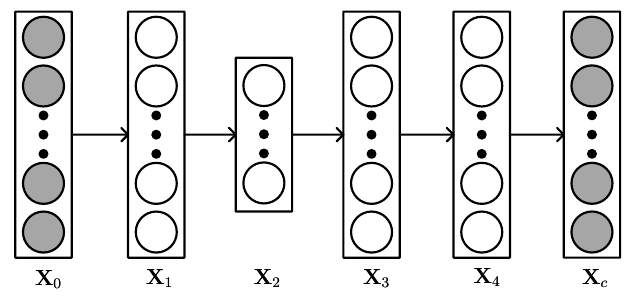}
    \caption{A 2-layer SDAE with $L=4$}
    \label{fig:sdae_model}
\end{figure}

SDAE \cite{conf_kdd_WangWY15} is an auto-encoder that learns the representation of the corrupted input by learning to predict its clean output. $\mathbf{X}_0$, the input matrix, is the noise-corrupted bag-of-words vectors of the texts, and $\mathbf{X}_L$ is the output, which is the clean version of the bag-of-words vectors. The architecture of the SDAE is shown in Fig.\ref{fig:sdae_model}. The generative process of layer \textit{l} in Bayesian SDAE is as follows:
\begin{enumerate}
  \item For each layer $l$, draw the weights and biases:
    \begin{align}
      W_l &\sim \mathcal{N}(0, \sigma_w^{2}\mathbf{I}_K)\\
      \mathbf{b}_l & \sim \mathcal{N}(0, \sigma_b^{2}\mathbf{I}_K)
    \end{align}
  \item For the output layer $L$, draw the clean versions:
  \begin{equation}
  \mathbf{X}_L\sim \mathcal{N}\Big(f_r(X_0, W), \sigma_n^{2}\mathbf{I}_K\Big)
  \end{equation}
\end{enumerate}
where $f_r(X_0, \mathbf{W})$ is the function that takes $X_0$ as the input and produces the constructed versions of the data.

The embedding vector in the middle layer, $\mathbf{X}_{L/2,i}$ corresponding to item $i$ is used as the textual representation of item $i$.

\section{Proposed Method}
\label{sec:methodology}
In this section, we describe our method, which is a combination of probabilistic matrix factorization for rating prediction and item embedding model for implicit feedback data. First, we will present the idea of item embedding model.

\subsection{Item Embedding from Click Data}
Inspired by word embedding techniques \cite{confnipsMikolovSCCD13,mikolov2013efficient,le2014distributed,conf/emnlp/LiZM15,levy2014neural} which represent a word by vectors that capture the relationship with its surrounding words, we apply the same idea to find representations of items based on implicit feedback data.

Similar to words, items also have their \textit{contexts}, which is a model choice and can be defined in different ways. For example, the context can be defined as the set of items that are clicked by a user (user-based context); or can be defined as the items that are clicked in a session with a given item (session-based context). In this work, we use the user-based context, which is defined as follow: given an item $i$ that is clicked by user $u$, the context of $i$ is the list of all items that $u$ have clicked.

The item embedding model presented in this section is partly based on the word-embedding model presented in \cite{conf/emnlp/LiZM15} which we bring into the world of items with a change: in \cite{conf/emnlp/LiZM15}, each word is represented by a unique vector while in this item embedding model we use two vectors to represent each item. We found that a model that uses two vectors for representing items can be efficiently trained by parallelizing the algorithm for the optimization problem (see our discussion on parameter learning in Section \ref{sec:generative_collaborative_item_embedding_model}).

Each item is characterized by two vectors: the \textit{feature vector} $\bm{\beta}_i$ and the \textit{context vector} $\bm{\alpha}_i$. These two vectors have different roles: the item feature vector describes the attributes of the item, while the context vector governs the distribution of the items that co-occur in its contexts.

The model describes the appearance of an item conditional on other items in its context as follows.
\begin{equation}
  \label{eq:embedding_equation}
  p(i|j)=f(i,j)p(i)
\end{equation}
where $p(i)$ is the probability that item $i$ appears in the data, $p(i|j)$ is the probability that $i$ appear in context of $j$ and $f(i,j)$ is the link function that reflects the association between $i$ and $j$. The role of the link function is straightforward: if item $i$ is often clicked (i.e., $p(i)$ is high), however, $i$ and $j$ are not often clicked together (i.e., $p(i|j)$ is low) then the link function $f(i,j)$ should have small value. On the other hand, if $i$ is rarely clicked (i.e., $p(i)$ is low) but if $i$ and $j$ are often clicked together (i.e., $p(i|j)$ is high), the link function $f(i,j)$ should have high value.

There are different choices for the link functions which lead to different embedding models. Following the work in \cite{conf/emnlp/LiZM15}, we choose the link function $f(i,j)=\exp\{\bm{\beta}_i^\top\bm{\alpha}_j\}$. Combine with Eq.\ref{eq:embedding_equation} we have:

\begin{equation}
\frac{p(i|j)}{p(i)}=\exp\{\bm{\beta}_i^\top\bm{\alpha}_j\}
\end{equation}

or:

\begin{equation}
  \label{eq:embedding_2}
  \log\frac{p(i|j)}{p(i)}=\bm{\beta}_i^\top\bm{\alpha}_j
\end{equation}

Note that $\log\frac{p(i|j)}{p(i)}=\log\frac{p(i,j)}{p(i)p(j)}$ is the point-wise mutual information (PMI) \cite{church90} of $i$ and $j$, Eq.(\ref{eq:embedding_2}) can be rewritten as follows.
\begin{equation}
  \label{eq:embedding_3}
  \bm{\beta}_i^\top\bm{\alpha}_j=PMI(i,j)
\end{equation}
Empirically, PMI can be estimated using the actual number of observations in the implicit feedback data.

\begin{equation}
\label{eq:empirical_pmi}
PMI(i,j)=\log\frac{\#(i,j)|\mathcal{D}|}{\#(i)\#(j)}
\end{equation}
where $\mathcal{D}$ is the set of all item-item pairs that are observed in the click history of any user, $\#(i)$ is the number of times item $i$ is clicked, $\#(j)$ is the number of times item $j$ is clicked, and $\#(i,j)$ is the number users who clicks both $i$ and $j$.

From Eq.\ref{eq:embedding_3} and Eq.\ref{eq:empirical_pmi} we can observe that, the item vectors and context vectors can be obtained by factorizing the matrix whose elements are defined in Eq.\ref{eq:empirical_pmi}.

A practical issue arises here: for item pair $(i, j)$ that are less often clicked by the same user, $PMI(i,j)$ is negative, or if they have never been clicked by the same user, $\#(i,j)=0$ and $PMI(i,j)=-\infty$. However, a negative value of PMI does not necessarily imply that the items are not related. The reason may be because the number of items is very huge, and a user who clicks $i$ may not know about the existence of $j$. A common way in natural language processing is to replace the negative values with zeros to form the positive PMI (PPMI) matrix \cite{bullinaria07}. The PPMI matrix $S$ whose elements are defined as follows.

\begin{equation}
  s_{ij}=\max\{PMI(i,j),0\}
\end{equation}

The item embedding model for implicit feedback can be summarized as follows: (1) construct an item-item matrix $S$ regarding the co-click of items in the click history of users (the PPMI matrix), and (2) factorize matrix $S$ to obtain the representations of items. The factorization can be performed following the PMF method presented in Section \ref{subsub:pmf}.

\subsection{Generative Collaborative Joint Model}
\label{sec:generative_collaborative_item_embedding_model}
We have presented the idea of item embedding via factorizing the PPMI matrix regarding items from the click data. We are ready to present the idea to combine the click model with the MF model for the rating data. Rather than performing two independent models: \textit{matrix factorization} on user ratings, and \textit{item embedding} on clicks, we connect them into a unified model which we describe below.

In item embedding above, item $i$ will be represented by two vectors: \textit{item feature vector} $\bm{\beta}_i$ and \textit{context vector} $\bm{\alpha}_i$ which are derived from the observations of the PPMI matrix. The item embedding vector $\bm{\beta}_i$ will be used as the item vector for the rating model.

The generative process of the joint model is as follows.

\begin{enumerate}
  \item Textual model: for each layer $l$, draw weight matrix $\mathbf{W}_l$, biases $\mathbf{b}_l$ and output $\mathbf{X}_l$
      \begin{align}
          \mathbf{W}_l \quad & \sim \quad \mathcal{N}(\mathbf{0}, \lambda_W^{-1}\mathbf{I}_K)\\
          \mathbf{b}_l \quad & \sim \quad \mathcal{N}(\mathbf{0}, \lambda_b^{-1}\mathbf{I}_K)\\
          \mathbf{X}_l \quad & \sim \quad \mathcal{N}(\mathbf{X}_{l-1}\mathbf{W}_l + \mathbf{b}_l), \lambda_x^{-1}\mathbf{I}_K)
      \end{align}
  \item For each item, draw feature vector $\bm{\beta}_i$, context vector $\bm{\alpha}_i$ and bias $\bm{\rho}_i$
    \begin{align}
      \bm{\beta}_i \quad & \sim \quad \mathcal{N}(\mathbf{X}_{L/2,i}, \sigma_\beta^2\mathbf{I}_K)\\
      \bm{\alpha}_i \quad & \sim \quad \mathcal{N}(0, \sigma_\alpha^2\mathbf{I}_K)
    \end{align}
  \item For each user $u$, draw user feature vector $\bm{\theta}_u$ and bias term $b_u$
  \begin{align}
      \bm{\theta}_u \quad & \sim \quad \mathcal{N}(\mathbf{0}, \sigma_\theta^2\mathbf{I}_K)
  \end{align}
  \item For each element $s_{ij}$ of matrix $S$
  \begin{align}
  s_{ij} \quad & \sim \quad \mathcal{N}(\bm{\beta}_i^\top\bm{\alpha}_j, \sigma^2_s)
  \end{align}
  \item For each pair $(u,i)\in\mathcal{R}$: draw the rating
    \begin{align}
      r_{uj} \quad & \sim \quad \mathcal{N}(\bm{\theta}_u^\top\bm{\beta}_i,\sigma^2_R)
    \end{align}
\end{enumerate}


\subsection{Parameter learning}
The model parameters are learned by maximizing the log posterior distribution which is equivalent to minimizing the following loss function.
\begin{equation}
  \label{eq:overal_objective_function}
  \begin{aligned}
\mathcal{L}\big(\bm{\Omega}\big) & = \frac{1}{2}\sum_{(u,i)\in \mathcal{R}}\Big(r_{ui}-\theta_i^\top\bm{\beta}_i\Big)^2 + \frac{\lambda_S}{2}\sum_{(i,j)\in \mathcal{S}}(s_{ij}-\bm{\beta}_i^\top\bm{\alpha}_j)^2\\
  \quad & +\frac{\lambda_\theta}{2}\sum_{u=1}^{N}||\bm{\theta}_u||^2_F +\frac{\lambda_\beta}{2}\sum_{i=1}^{M}||\bm{\beta}_i-f_e(\mathbf{X}_{0,i},\mathbf{W})^\top||^2_F +\frac{\lambda_\alpha}{2}\sum_{j=1}^{M}||\bm{\alpha}_j||^2_F\\
  \quad &+\frac{\lambda_X}{2}\sum_{i=1}^{M}||\mathbf{X}_{c,i}-f_r(\mathbf{X}_{0,i},\mathbf{W})^\top||^2_F +\frac{\lambda_W}{2}\sum_l\Big(||\mathbf{W}_l||^2_F+||\mathbf{b}_l||^2_F\Big)
  \end{aligned}
\end{equation}
where $\mathcal{S}=\{(i,j)|s_{ij}>0\},\lambda_S=\sigma_R^2/\sigma_S^2,\lambda_\theta=\sigma_R^2/\sigma_\theta^2$, $\lambda_\beta=\sigma_R^2/\sigma_\beta^2$, $\lambda_\alpha=\sigma_R^2/\sigma_\alpha^2$; $f_e(\mathbf{X}_{0,i},\mathbf{W})$ is the middle layer of the SDAE; $f_r(\mathbf{X}_{0,i},\mathbf{W})$ is the output layer of the SDAE.


For each user $u$, at each iteration, we calculate the partial derivative of $\mathcal{L}(\bm{\theta}, \bm{\beta}, \mathbf{W}, \mathbf{B})$ with respect to $\bm{\theta}_u$ while fixing other parameters. By setting this derivative to be zero: $\frac{\partial \mathcal{L}}{\partial \bm{\theta}_u}=0$, we obtain the update formula for $\bm{\theta}_u$ as:

\begin{equation}
  \label{eq:update_rule_theta}
  \bm{\theta}_u=\Big(\sum_{i\in \mathcal{R}_u} \bm{\beta}_i\bm{\beta}_i^\top+\lambda_\theta\mathbf{I}_K\Big)^{-1}\Big(\sum_{i\in \mathcal{R}_u} r_{ui}\bm{\beta}_i\Big)
\end{equation}

Similarly, we can respectively obtain the update formulas for $\bm{\beta}_i$, $\bm{\alpha}_i$, $\mathbf{W}$ and $\mathbf{B}$ using the same way.

\begin{equation}
  \label{eq:update_rule_item}
  \begin{aligned}
  \bm{\beta}_i =&\Big(\sum_{u\in \mathcal{R}_i} \bm{\theta}_u\bm{\theta}_u^\top+\lambda_S\sum_{j\in\mathcal{S}_i}\bm{\alpha}_j\bm{\alpha}_j^\top+\lambda_\beta\mathbf{I}_K\Big)^{-1}\\
  &\times \Big(\sum_{u\in \mathcal{R}_i} r_{ui}\bm{\theta}_u+\lambda_S\sum_{j\in\mathcal{S}_i}s_{ij}\bm{\alpha}_j+\lambda_\beta f_e\big(\mathbf{X}_{0,i},\mathbf{W}\big)\Big)
  \end{aligned}
\end{equation}
\begin{equation}
  \bm{\alpha}_j  =\Big(\lambda_S\sum_{i\in \mathcal{S}_j} \bm{\beta}_i\bm{\beta}_i^\top+\lambda_\alpha\mathbf{I}_K\Big)^{-1}\Big(\lambda_S\sum_{i\in \mathcal{S}_j} s_{ij}\bm{\beta}_i\Big)
\end{equation}
where $\mathcal{S}_j=\{i|s_{ij}>0\}$, $\mathcal{R}_u$, again, is the set of items that $u$ gave ratings, and $\mathcal{R}_i$ is the set of users that gave ratings to $i$.

Given $\bm{\theta,\beta,\alpha}$, we can learn the weights $\mathbf{W}_l$ and biases $\mathbf{b}_l$ for each layer $l$ of the SDAE, using the back-propagation learning algorithm. The gradients of the objective function with respect to $\mathbf{W}_l$ and $\mathbf{b}_l$ are given below.
\begin{equation}
\begin{aligned}
&\frac{\partial \mathcal{L}}{\mathbf{W}_l}=-\lambda_W\mathbf{W}_l\\
&-\lambda_\beta\sum_i \Big[\frac{\partial f_e\big(\mathbf{X}_{0,i},\mathbf{W}\big)}{\mathbf{W}_l}\Big]^\top\Big[f_e\big(\mathbf{X}_{0,i},\mathbf{W}\big)^\top-\bm{\beta}_i\Big]\\
&-\lambda_X\sum_i \Big[\frac{\partial f_r\big(\mathbf{X}_{0,i},\mathbf{W}\big)}{\mathbf{W}_l}\Big]^\top\Big[f_r\big(\mathbf{X}_{0,i},\mathbf{W}\big)^\top-\mathbf{X}_{c,i}\Big]
\end{aligned}
\end{equation}

\begin{equation}
\begin{aligned}
&\frac{\partial \mathcal{L}}{\mathbf{b}_l}=-\lambda_W\mathbf{b}_l\\
&-\lambda_\beta\sum_i \Big[\frac{\partial f_e\big(\mathbf{X}_{0,i},\mathbf{W}\big)}{\mathbf{b}_l}\Big]^\top\Big[f_e\big(\mathbf{X}_{0,i},\mathbf{W}\big)^\top-\bm{\beta}_i\Big]\\
&-\lambda_X\sum_i \Big[\frac{\partial f_r\big(\mathbf{X}_{0,i},\mathbf{W}\big)}{\mathbf{b}_l}\Big]^\top\Big[f_r\big(\mathbf{X}_{0,i},\mathbf{W}\big)^\top-\mathbf{X}_{c,i}\Big]
\end{aligned}
\end{equation}

By alternatively updating $\bm{\theta,\beta,\alpha}, \mathbf{W}_l$ and $\mathbf{b}_l$, we can find a local optimum for $\mathcal{L}$.

\textbf{Computational complexity.} For user vectors, as analyzed in \cite{hu2008collaborative}, the complexity for updating $N$ users in an iteration is $\mathcal{O}(K^2\mathcal{|R|}+K^3N)$, where $|\mathcal{R}|$ is the number of non-zero entries of rating matrix $R$. Since $|\mathcal{R}|>>N$,  if $K$ is small, this complexity is linear in the size of the input matrix. For item vector updating, we can easily show that the complexity for updating $M$ items in an iteration is $\mathcal{O}(K^2(|\mathcal{R}|+|\mathcal{S}|)+K^3M+VK_1M)$, where $\mathcal{|S|}$ is the number of non-zero entries of matrix $S$, $V$ is the size of the vocabulary, $K_1$ is the dimensionality of the first layer of the SDAE. Note that the term $\mathcal{O}(VK_1)$ is the cost for computing the output of the encoder, which is dominated by the computation of the first layer.

Therefore, the computational complexity for one epoch is $\mathcal{O}\Big(K^2(2|\mathcal{R}|+|\mathcal{S}|)+K^3(N+M)+VK_1M\Big)$. We can see that the computational complexity linearly scales with the number of users and the number of items. Furthermore, this algorithm is easy to be parallelized to adapt to large-scale data. For example, in updating user vectors $\boldsymbol\uptheta$, the update rule of user $u$ is independent of other users' vectors, therefore, we can compute $\sum_i\bm{\beta}_i\bm{\beta}_i^\top$ in advance, and update $\bm{\theta}_u$ in parallel.

\subsection{Rating prediction}
After learning parameters $\bm{\theta},\bm{\beta},\bm{\alpha},\mathbf{W},\mathbf{B}$, the proposed model can be used for predicting missing ratings. We consider two cases of rating predictions: \textbf{in-matrix} prediction and \textbf{out-of-matrix} prediction. In-matrix prediction refers to the case that we predict the rating of user $u$ to item $i$, where $i$ has not been rated by $u$ but has been rated by at least one other users. Out-matrix refers to the case that we predict the rating of user $u$ to item $i$, where $i$ has not been rated by any users (i.e., $i$ has implicit feedback only).

Let $\mathcal{R}$ be the observed data (observed rating scores), the unobserved $r_{ui}$ can be estimated as follows.

In-matrix prediction: $\mathbb{E}\big[r_{ui}|\mathcal{R}\big]=\bm{\theta}_u^\top\bm{\beta}_i$

Out-of-matrix prediction: $\mathbb{E}\big[r_{ui}|\mathcal{R}\big]=\bm{\theta}_u^\top f_e(\mathbf{x}_i,\mathbf{W})$

\section{Empirical study}
\label{sec:empirical_study}

\subsection{Datasets}
We use two public datasets in different domains. The datasets are:

  \textbf{MovieTweetings}: a dataset of user-movie ratings collected from via Twitter users. It contains 630,000 ratings in the range 1-10 to 28,000 movies of 50,000 users. We crawled movie plot summaries from the website \url{https://www.imdb.com/}, an online database of information related to movies, and use as movies' textual information.

  \textbf{Bookcrossing}: A dataset collected by Cai-Nicolas Ziegler in August and September 2004 from the Book-Crossing\footnote{http://www.bookcrossing.com/}. The dataset contains 278,858 users (anonymized but with demographic information) providing 1,149,780 ratings (explicit/implicit) about 271,379 books. We remove users and items that have no explicit feedback. We use the book descriptions crawled from \url{http://www.lookupbyisbn.com} as textual information of the books.

The statistical information about the datasets is given in Table \ref{tab:data_information}.

\begin{table}[!t]
  \centering
  \caption{Statistical information of the datasets}
  \label{tab:data_information}
  \begin{tabular}{ccc}
    \toprule
    &MovieTweetings&Bookcrossing\\
    \midrule
    \hline
    \# of users & 50,000& 77,805\\
    \# of items & 28,000& 185,973\\
    value of ratings & 1 -- 10 & 1--10\\
    average rating & 3.53 & 7.61\\
    \# of ratings & 630,000 & 357,246\\
    rating density (\%) & 0.045 & 0.0029\\
    \# of clicks & - & 892,185\\
    click density (\%) & - & 0.0062 \\
  \bottomrule
\end{tabular}
\end{table}

Since MovieTweetings datasets contain only explicit feedback, we artificially create the implicit feedback and explicit feedback data following \cite{DBLP:conf/icdm/BellK07}. For the implicit feedback, we use all the rating data by considering whether a user rated an item or not. In other words, the implicit feedback is obtained by binarizing the rating data. For explicit feedback, we randomly pick 10\%, 20\%, 50\%, and 80\%, from the rating data and use as explicit feedback. Details of datasets obtained are given in Table \ref{tab:ml_subsets_mt}.

\begin{table}
\centering
\caption{Datasets obtained by picking ratings from the \textit{MovieTweetings}}
\label{tab:ml_subsets_mt}
\setlength{\tabcolsep}{0.4em}
\begin{tabular}{c|c|c}
    \hline
    Dataset &\thead{\% rating\\picked} & \thead{Density of the\\rating matrix (\%)}\\
    \hline
    MT-10 & 10\% & 0.3561\\
    MT-20 & 20\% & 0.6675\\
    MT-50 & 50\% & 1.6022\\
    MT-80 & 80\% & 1.9348\\
    \hline
\end{tabular}
\end{table}

\subsection{Evaluation}
We split the rating data into two parts: 80\% for the training set and 20\% for as ground-truth for testing. From the training set, we randomly pick 10\% as a validation set that will be used for model selection and checking stopping condition of the training phase. In evaluating the in-matrix prediction, when splitting data, we make sure that all the items in the test set appear in the training set (to ensure that all the items in the test set have at least one rating in the past). In evaluating out-of-matrix prediction, we make sure that none of the items in the test set appear in the training set (to ensure that none of the items in the test set have any rating in the past).

The model is trained on the training dataset and the optimal parameters are obtained by using the validation set. The model with these optimal parameters is then used to predict ratings for user-item pairs that appear in the test set. We use Root Mean Square Error (RMSE), as the metric to measure the performance of the models. RMSE measures the deviation between the rating predicted by the model and the true ratings (given by the test set) and is defined as follows.
\begin{equation}
\label{eq:RMSE}
RMSE=\sqrt{\frac{1}{|Test|}\sum_{(u,i)\in Test}(r_{ui}-\hat{r}_{ui})^2}
\end{equation}
where $|Test|$ is the size of the test set. The smaller the value of RMSE on the test set is, the better the performance of the model is.

\subsection{Competing methods}
\textbf{For in-matrix prediction}. We compare our method with three factorization models as follows.
\begin{enumerate}
\item \textit{PMF} \cite{salakhutdinov2008a}: a state-of-the-art method for rating prediction which we described in Section \ref{sec:preliminary}.
\item \textit{CTR} \cite{wang2011collaborative}: Collaborative Topic Regression is a state-of-the art recommendation model, which combines collaborative filtering (PMF) and topic modeling (LDA) to use both ratings and documents.
\item \textit{CDL} \cite{conf_kdd_WangWY15}: Collaborative Deep Learning is another state-of-the-art recommendation model, which enhances rating prediction accuracy by analyzing documents using stacked denoising auto encoder (SDAE).
\item \textit{CVAE} \cite{Li_CVAE_2017}: Collaborative Variational AutoEncoder is another state-of-the-art recommendation model, which uses Variational AutoEncoder for modeling the texts.
\item \textit{TCF}: Textual Co-Factorization is the proposed model of this paper.
\end{enumerate}

\subsection{Parameter settings}
In all settings, we set the dimension of the latent space to $K=64$. For PMF, CTR, CDL and CVAE, we used the grid search to find the optimal values of the regularization terms that produce the best performance on the validation set. For our proposed method, we explored different settings of hyper-parameters to study the influence of the hyper-parameters to the performance of the model.

\subsection{Results}

We report the RMSE on the test set for in-matrix prediction in Table \ref{tab:in_matrix_comparision} and out-of-matrix prediction in Table \ref{tab:out_matrix_comparision}.

\textbf{Comparison over methods.} From the results, we have following observations.
\begin{itemize}
  \item The hybrid models which use textual contents (CTR, CDL, CVAE, TCF) outperform the PMF which uses only the interaction data. Specially, when the data is very sparse (the MT-10), the differences between PMF and the other competitors are most significant. This indicates the benefit of exploiting textual information in rating prediction, particularly, for extremely sparse data.
  \item In all datasets, TCF outperforms CTR, CDL and CVAE. It indicates that introducing click data will improve the accuracy of the model. Especially, for when the data becomes more sparse (from MT-80 to MT-20), the differences between TCF and other methods increases. This indicates that, for sparse datasets, introducing the click data will improve the accuracy of the model.
  \item In the out-of-matrix prediction (Table \ref{tab:out_matrix_comparision}), only CTR, CDL, CVAE and CTF work. All the three methods perform worse than themselves in the in-matrix prediction. This is reasonable because the prior ratings are not involved in the prediction. CTF performs better than CTR, CDL, and CVAE, which indicates the benefit of utilizing click information.
\end{itemize}
\begin{table}[t]
    \caption{The RMSE for the in-matrix prediction on MT-50 and Bookcrossing datasets}
    \centering
    \begin{subtable}{1.0\linewidth}
      \centering
      \caption{In-matrix prediction}
      \label{tab:in_matrix_comparision}
      \setlength{\tabcolsep}{0.4em}
      \begin{tabular}{l|c|c}
          \hline
          Methods &MT-50 & Bookcrossing\\
          \hline
          PMF & 1.4685 & 2.1663\\
          CTR & 1.4538 & 1.5982\\
          CDL & 1.4412 & 1.5041\\
          CVAE & 1.3729 & 1.4827\\
          TCF (our) & 1.1637 & 1.3133\\
          \hline
      \end{tabular}
    \end{subtable}%
    \bigskip
    \hfill
    \begin{subtable}{1.0\linewidth}
      \centering
      \caption{Out-of-matrix prediction}
      \label{tab:out_matrix_comparision}
      \setlength{\tabcolsep}{0.4em}
      \begin{tabular}{l|c|c}
          \hline
          Methods &MT-50 & Bookcrossing\\
          \hline
          CTR & 1.6985 & 1.6971\\
          CDL & 1.6231 & 1.6126\\
          CVAE & 1.5824 & 1.5945\\
          TCF (our) & 1.3412 & 1.4981\\
          \hline
      \end{tabular}
    \end{subtable}
\end{table}

\begin{figure}
    \centering
    \begin{subfigure}[b]{0.5\textwidth}
        \includegraphics[width=\textwidth]{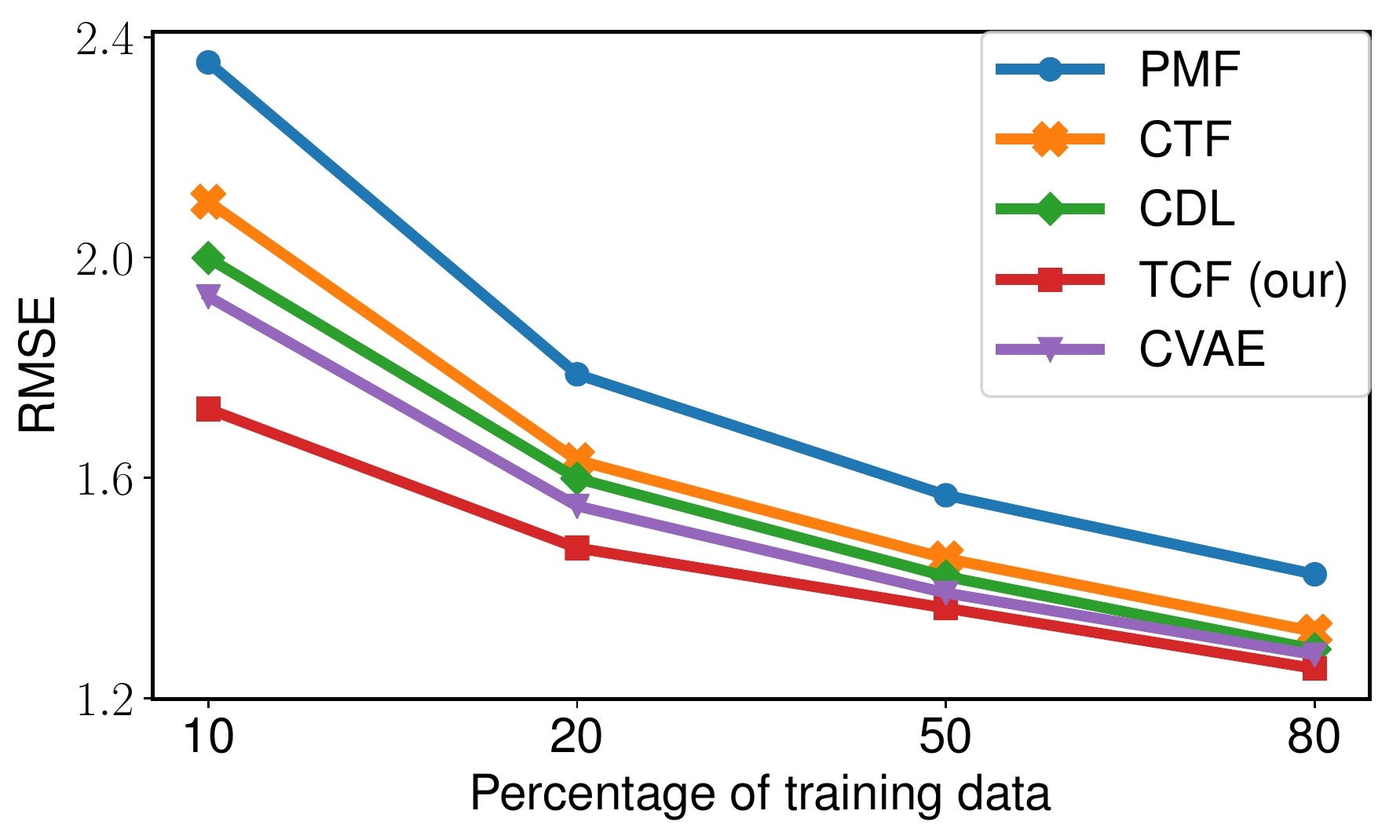}
        \caption{In-matrix prediction}
        \label{fig:in_matrix_sparse}
    \end{subfigure}
    ~ 
    \begin{subfigure}[b]{0.5\textwidth}
        \includegraphics[width=\textwidth]{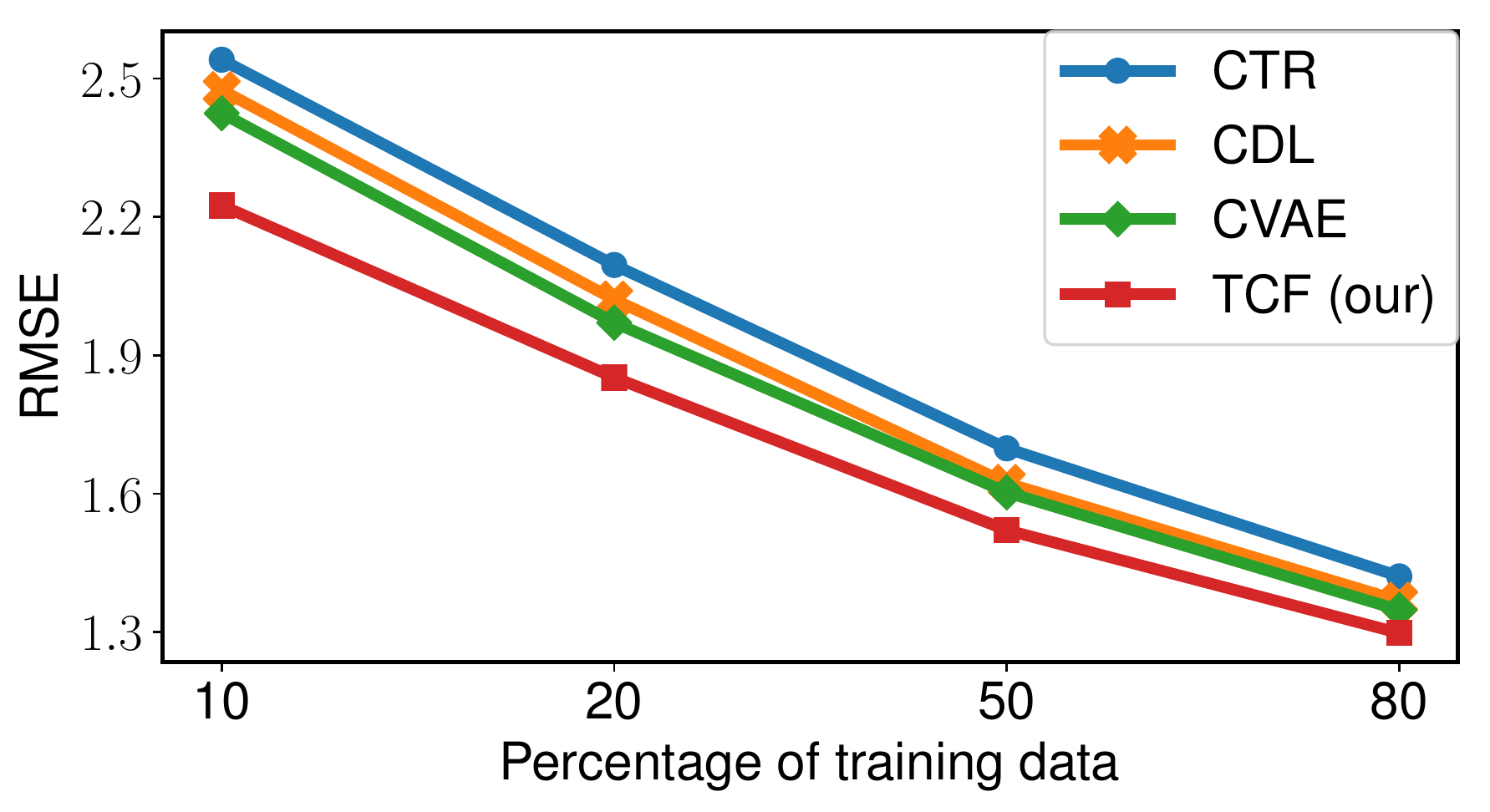}
        \caption{Out-of-matrix prediction}
        \label{fig:out_matrix_sparse}
    \end{subfigure}
    \caption{The RMSE for the data of different levels of sparsity}
    \label{fig:sparsity_accuracy}
\end{figure}

\textbf{Performances on sparse datasets.} We study the performances of the models on different levels of the sparsity of the rating data. Fig.\ref{fig:in_matrix_sparse} and Fig.\ref{fig:out_matrix_sparse} shows the results of in-matrix and out-of-matrix prediction over subsets of the MovieTweetings data.

We can observe that:
\begin{itemize}
  \item For all methods, the accuracies increase with respect to the densities of the rating data. This is expected because we have more training data for learning good representations.
  \item In all cases, the proposed method TCF outperforms CTR and CDL, indicating that the click data help in improving the accuracy of rating prediction.
  \item The differences between TCF and other methods are more pronounced in the most sparse data MT-10. This indicates that the click data has a significant role in sparse datasets.
\end{itemize}

\begin{figure}
    \centering
    \includegraphics[width=0.5\textwidth]{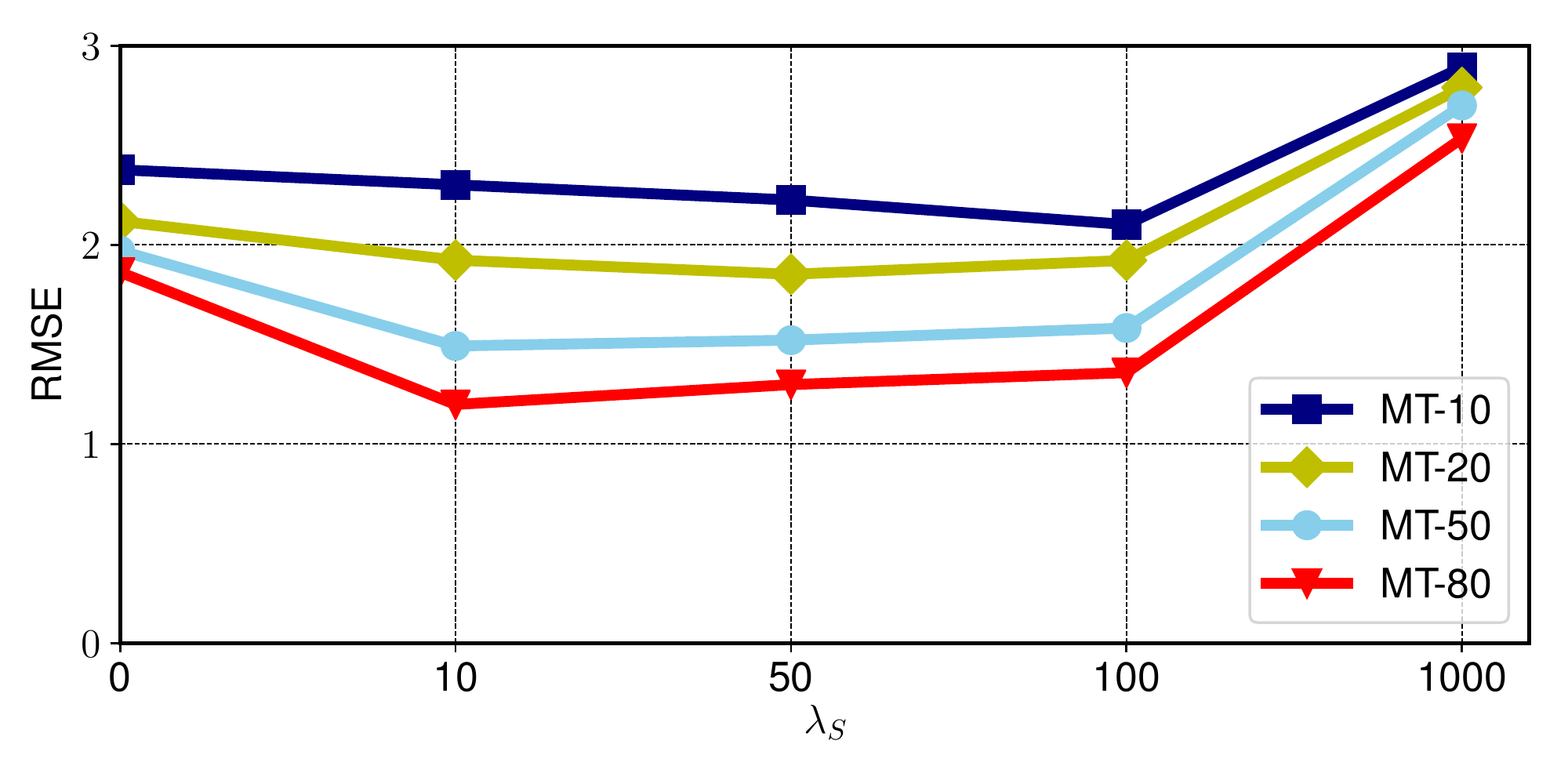}
    \caption{Test RMSE of in-matrix prediction task on different subsets of MovieTweetings dataset corresponding to different values of $\lambda_S$.}
    \label{fig:in_matrix_prediction_varying_lambda}
\end{figure}

\textbf{Impact of the parameter } $\lambda_S$. As in the Eq.\ref{eq:overal_objective_function}, parameter $\lambda$ controls the level of contribution of implicit feedback data to the model. If $\lambda_S=0$, the model reduces to the original CDL which uses explicit feedback data only for modeling users and items. If $\lambda_S=\infty$, the model uses only information from the implicit feedback to model items. In this part, we vary $\lambda_S$ while fixing other parameters to study the effect of $\lambda_S$ on the accuracy of the model.
Figure \ref{fig:in_matrix_prediction_varying_lambda} shows the test RMSE of in-matrix prediction task of our proposed method when the $\lambda_S$ is varied.

From the result, we can observe that the prediction performance is influenced significantly by the value of $\lambda_S$. For small values of $\lambda_S$, the test RMSE is relatively high, it decreases when $\lambda_S$ increases. However, when $\lambda_S$ goes over a certain threshold, the test RMSE starts increasing. This can be explained as follows. For a very small value of $\lambda_S$, the model mainly uses information from the explicit feedback which is too sparse to model the users and items. When the value of $\lambda_S$ becomes very large, the model mainly uses the implicit feedback data for modeling the items, therefore, is not reliable. The best values of $\lambda_S$ should balance the contribution of implicit and explicit feedback.

\section{Related Work}
\label{sec:related_work}
Collaborative filtering \cite{salakhutdinov2008a,nguyen2018npe,nguyen2015city,nguyen2017probabilistic,nguyen2017collaborative,nguyen2017hierarchical,hu2008collaborative}suffers from the cold-start problem. To overcome the problem, exploiting the auxiliary data such as textual data \cite{wang2011collaborative,conf_kdd_WangWY15,Li_CVAE_2017,Cheng_MMALFM,nguyen2019learning} and visual data \cite{conf/aaai/HeM16}.

An approach for addressing the cold-start problem is to utilize the click data, which is much easier to collect than the rating data. Co-rating \cite{conf/cikm/LiuXZY10} combines explicit (rating) and implicit (click) feedback by treating explicit feedback as a special kind of implicit feedback. The explicit feedback is normalized into the range $[0, 1]$ and is summed with the implicit feedback matrix with a fixed proportion to form a single matrix. This matrix is then factorized to obtain the latent vectors of users and items.

Wang et. al.\cite{conf/pakdd/WangRZW12} proposed Expectation-Maximization Collaborative Filtering (EMCF) which exploits both implicit and explicit feedback for the recommendation. For predicting ratings for an item, which does not have any previous ratings, the ratings are inferred from the ratings of its neighbors according to click data.

The main difference between these methods with ours is that they do not have a mechanism for balancing the amounts of click data and rating data when making predictions. In our model, these amounts are controlled depending on the number of previous ratings that the target items have.

In \cite{Pan:2019:TRH:3289475.3243652}, the author proposed a transfer learning-based model for multiple data sources in collaborative filtering (the browsing data and the purchase data). This model solves the recommendation with multiple implicit feedbacks. Different from this work, we address the rating prediction (explicit feedback) by exploiting the click data and textual contents.

Item2Vec \cite{confrecsysBarkanK16} is a neural network-based model for learning item embedding vectors using \textit{co-click} information. In \cite{confrecsysLiangACB16}, the authors applied a word embedding technique by factorizing the shifted PPMI matrix \cite{levy2014neural}, to learn item embedding vectors from click data. However, using these vectors directly for rating prediction is not appropriate because click data does not exactly reflect the preferences of users. Instead, we combine item embedding with MF in a way that allows rating data to contribute to item representations.

\section{Conclusion}
\label{sec:conclusion}
In this paper, we proposed a probabilistic model that exploits click data for addressing the cold-start problem in rating prediction. The model is a combination of two models: (i) an item embedding model for click data, and (ii) MF for rating prediction. The experimental results showed that our proposed method is effective in rating prediction for items with no previous ratings and also boosts the accuracy of rating prediction for extremely sparse data.

\section*{Acknowledgments.} This work was supported by a JSPS Grant-in-Aid for Scientific Research (B) (15H02789, 15H02703).

\bibliographystyle{unsrt}
\bibliography{tcf_journal}

\end{document}